\documentclass[twocolumn,aps,prl,showpacs]{revtex4}
\input epsf
\input{epsf.sty}

\begin{document}

\hsize\textwidth\columnwidth\hsize
\csname@twocolumnfalse\endcsname
 
\title{ Vortex fluctuations in underdoped
Bi$_{2}$Sr$_{2}$CaCu$_{2}$O$_{8+\delta}$ crystals}

\author {Sylvain Colson,$^{1}$  Marcin Konczykowski,$^{1}$
Marat B. Gaifullin,$^{2}$ Yuji Matsuda,$^{2}$ Piotr Gier\l owski,$^{3}$ Ming 
Li,$^{4}$ \\ Peter H. Kes$^{4}$ and Cornelis J. van der Beek,$^{1}$}
\affiliation{$^{1}$Laboratoire des Solides Irradi\'{e}s, 
CNRS-UMR 7642 and CEA/DSM/DRECAM, Ecole Polytechnique, 91128  Palaiseau, 
France\\
$^{2}$Institute for Solid State Physics, University 
of Tokyo, Kashiwanoha, Kashiwa, Chiba 277-8581, Japan\\
$^{3}$Institute of Physics, Polish Academy of 
Sciences, Al. Lotnik\'{o}w 32/46, 02-668 Warsaw, Poland\\
$^{4}$ \mbox{Kamerlingh~Onnes~Laboratorium,~Leiden~University,~P.O.~Box~9506,~2300~RA~Leiden,~the~Netherlands}
}

\date{\today}

\begin{abstract}
\noindent
 Vortex thermal fluctuations in heavily underdoped Bi$_{2}$Sr$_{2}$CaCu$_{2}$O$_{8+\delta}$
($T_{c}$=69.4~K) are studied using Josephson plasma resonance (JPR). From 
the zero-field data, we obtain the $c$-axis penetration depth 
$\lambda_{L,c}(0)Ê=Ê230 \pm 10$ $\mu$m and the 
anisotropy ratio $\gamma(T)$. The low plasma frequency allows us to study phase correlations over 
the whole vortex solid state, and to extract a wandering length $r_{w}$  of vortex pancakes.
The temperature dependence of $r_{w}$ as well as its increase with dc magnetic field 
is explained by the renormalization of the vortex line tension by the
fluctuations, suggesting that this softening is responsible for the dissociation of the 
vortices at the first order transition.

\end{abstract}


\maketitle


Vortex thermal fluctuations are considered 
essential in determining the $(H,T)$ phase diagram of layered high 
temperature superconductors (HTS), and notably the first order transition (FOT)
in which the ordered vortex crystal transforms to a liquid state 
without long range phase coherence \cite{Cubitt93,Zeldov95}.  Many scenarios, 
varying from vortex lattice melting described by a Lindemann criterion 
\cite{Blatter96} to layer decoupling 
\cite{Koshelev91,Daemen93,Dodgson00},  all 
considering various degrees of coupling between the  superconducting layers,
have successfully been used to describe the position of the FOT in 
the $(H,T)$--plane. However, such fits to the FOT  line
have not been able to convincingly discriminate between the different 
models. Here, we aim to do just this, through a direct measurement of 
the amplitude, as well as the field and temperature dependence of vortex thermal 
excursions in the vortex solid phase (or ``Bragg--glass'' \cite{Giamarchi97})
that lead to the FOT. 

For this study, we use the layered 
Bi$_{2}$Sr$_{2}$CaCu$_{2}$O$_{8+\delta}$ (BSCCO) compound, in which 
vortex excursions can conveniently be measured by the Josephson Plasma 
Resonance (JPR) technique \cite{Tsui94,Gaifullin00,Shibauchi99}.
Briefly, the interlayer Josephson current $J^{(c)}_{m}$ can be 
measured through the JPR frequency $\omega_{pl}\sim J^{(c) 1/2}_{m}$, at 
which the equality of charging and kinetic energy leads to a 
collective excitation of Cooper pairs across the layers. In turn, 
$\omega_{pl}^{2}(B,T) = \omega_{pl}^{2}(0,T)\langle \cos(\phi_{n,n+1})\rangle$
intimately depends on the gauge-invariant phase difference $\phi_{n,n+1}$
between adjacent layers $n$ and $n+1$  \cite{Bulaevskii95}.
Here, $\langle\ldots\rangle$ stands for thermal and disorder 
averaging. Thus, JPR is a probe of the interlayer phase coherence. 
The fluctuations of vortex lines created by a dc magnetic field applied 
perpendicularly to the layers modify the relative phase difference between adjacent 
layers and thus depress $\omega_{pl}$. 
In Bi$_{2}$Sr$_{2}$CaCu$_{2}$O$_{8+\delta}$, the ensemble of vortex lines should be 
described as stacks of two-dimensional pancake vortices. 
Thermal fluctuations 
shift the individual vortex pancakes with respect to their nearest 
neighbors in the $c$ direction, by a distance ${\bf r}_{n,n+1}= {\bf u}_{n+1} - {\bf u}_{n}$. 
Here {\bf u}$_{n}$ is the $ab$-plane displacement of the pancake 
vortex in layer~$n$ with respect to the equilibrium position of the 
stack it belongs to (Fig.~1).  The wandering length of vortex lines, which is related directly to the JPR 
frequency $\omega_{pl}^{2}$, can be then defined as $r_{w} = 
{\langle {\bf r}_{n,n+1}^{2} \rangle}^{1/2}$ \cite{Bulaevskii00,Koshelev00}.
Below, we shall only consider temperatures above $T = 42$ K, at which
vortex pinning (quenched disorder) is unimportant \cite{Zeldov95,Avraham01}.


\begin{figure}[t]
\centerline{\epsfxsize 3.2cm \epsfbox{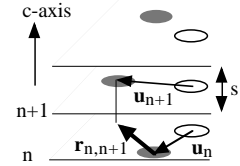}}
\caption{Meandering of vortex pancakes along the vortex line in 
layered superconductors. Because of thermal fluctuations
and disorder, pancakes (full circles) are shifted away from
their equilibrium position (open circles).}
\end{figure}

Underdoped BSCCO ($T_{c}$ = 69.4~K) single crystals were grown by the traveling solvent
floating zone method in 25~mbar O$_{2}$ partial pressure at the FOM-ALMOS center, the Netherlands \cite{MingLi02}.
The samples were post-annealed for one week at 700$^{\circ}$C in 
flowing N$_{2}$. The advantage of using heavily underdoped BSCCO is that 
$\omega_{pl}(0,0)\approx$~61 GHz turns out to be very low, 
which allows us to measure the vortex meandering over the entire vortex phase diagram.
Samples A and B (cut from the same crystal) have  dimensions $1.35 \times 
1 \times 0.04$~mm$^{3}$ and $0.7 \times 0.47 \times$$ 0.04$~mm$^{3}$, respectively. 
Another sample from the same batch was used to determine the temperature of the~FOT
(Fig.~\ref{fig:rw-vs-t}b, inset), by measuring the paramagnetic
peak at the FOT with a miniature Hall probe magnetometer \cite{Morozov96}.

\begin{figure}[h]
		\centerline{\epsfxsize 8cm \epsfbox{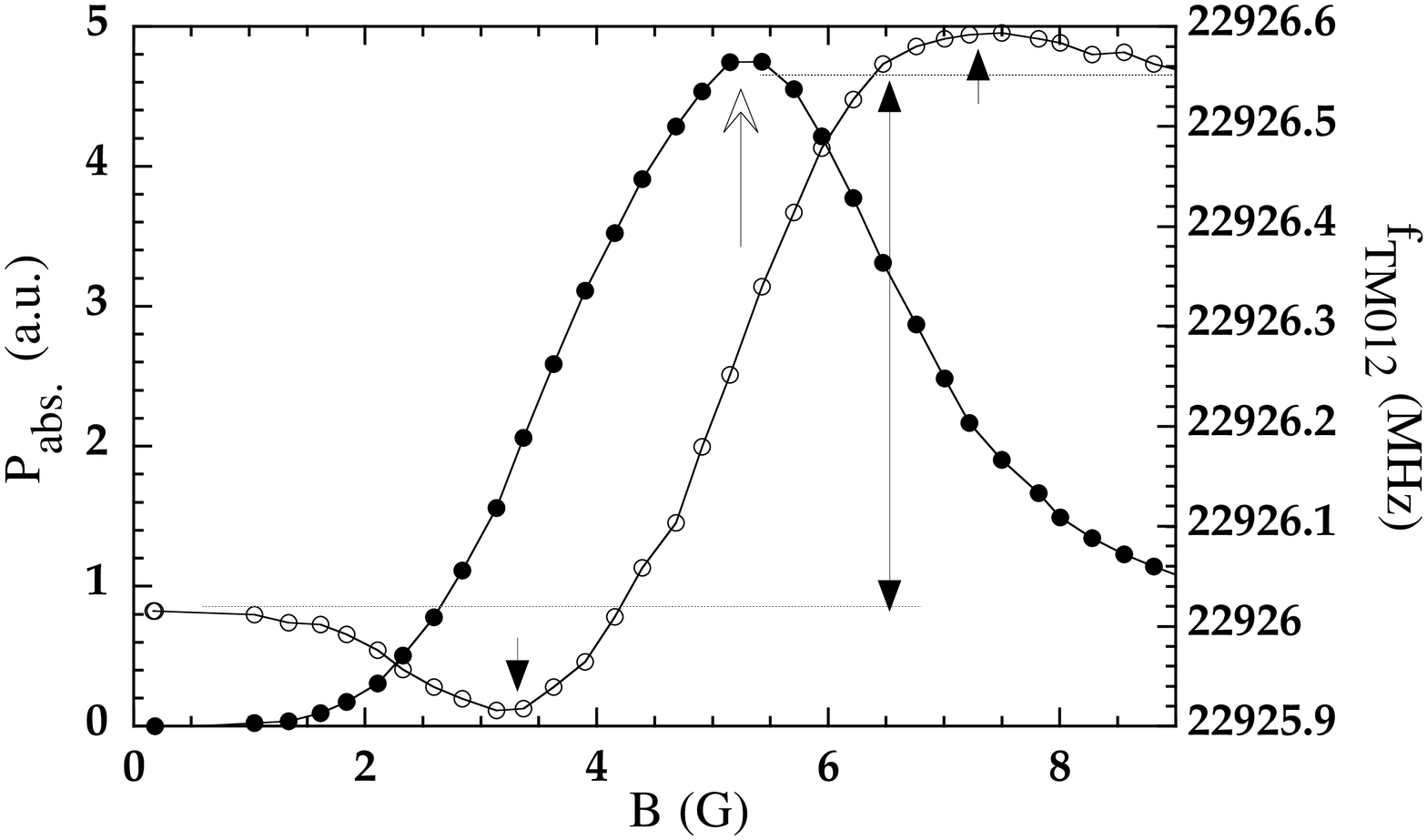}}
		\vspace{-3mm}
		\caption{Field sweep experiment on sample A at $T= 66$~K 
		in the TM$_{012}$ mode of the cavity ($f=22.9$ GHz). The power absorbed 
		($\bullet$, left axis) in the sample 
		shows a maximum at $B_{JPR}$$ = 5.3$ G at which 
		$\omega=\omega_{pl}$ (open arrow). 
		At the same field, the resonance frequency of the cavity ($\circ$, 
		right axis) displays a double-peak structure, indicated by the 
		closed arrows, and a jump (arrow between dashed lines).}
\end{figure}

The JPR measurements were
carried out using  the cavity perturbation technique in 
the Laboratoire des Solides Irradi\'{e}s (on sample A) and the 
bolometric method in the Institute for Solid State Physics 
at the University of Tokyo (on samples A and B). For the cavity 
perturbation technique, the sample was glued in the center of the top 
cover of a cylindrical Cu cavity used in the different 
 TM$_{01i}$ ($i= 0,\ldots,4$) modes. These provide the correct 
configuration of the microwave field at the sample location, in which 
$E_{rf}{\parallel}c$-axis and $H_{rf}\approx 0$ \cite{Colson02}. The
 unloaded quality factor $Q_{0}$ is measured as function of 
 temperature and field to obtain the power absorbed by the sample (Fig.~2).
The bolometric method \cite{Matsuda94} consists in measuring the 
heating of the sample induced by the absorption of the incident microwave 
power when the JPR is excited~\cite{Gaifullin00,Colson02}. 
Furthermore, reversible magnetization measurements were carried out 
on sample A using a 
commercial superconducting quantum interference device magnetometer in order to extract 
the London penetration depth  $\lambda_{L,ab}(T)$ for currents in the $ab$--plane
\cite{MingLi02}.

Figure 3 shows the JPR frequency $f_{JPR} = \omega_{pl}/2\pi$ in zero field obtained 
by the above-mentioned methods on samples A and B. 
$\omega_{pl}^{2}$ is proportional to 
the maximum interlayer Josephson current along the
$c$-axis \cite{Bulaevskii95},

\begin{equation}
	\omega_{pl}^{2}(H,T) = \omega_{pl}^{2}(0,T) \langle 
	\cos(\phi_{n,n+1})\rangle = \frac {2\pi\mu_{0}c^{2}s}{\epsilon_{r}\Phi_{0}} 
	J^{(c)}_{m}(B,T)
	\label{eq:omega plasma}
\end{equation}

\begin{figure}[h]
		\centerline{\epsfxsize 8cm \epsfbox{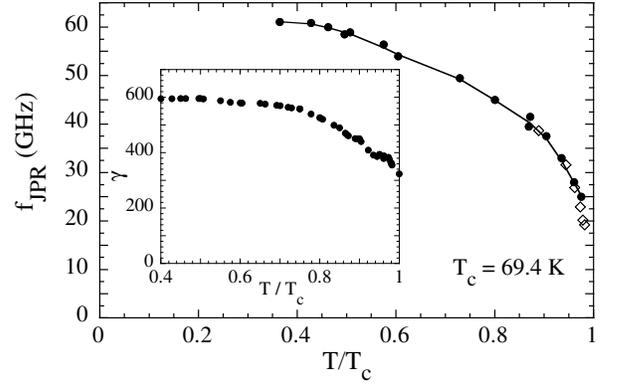}}
		\vspace{-3mm}
		\caption{ The JPR frequency in zero magnetic field for samples 
		A (data below 40 GHz) and B (data above 40 GHz) 
		vs the reduced temperature $T/T_{c}$. The bolometric method 
		($\bullet$) and the cavity perturbation technique ($\diamond$) 
		have been used. We use a spline fit (solid line) in the 
		extraction of the wandering length (see text). Inset: experimental 
		temperature dependence of $\gamma$, obtained by the division of 
		the experimental $\lambda_{L,c}(T)$ data by the 
		$\lambda_{L,ab}(T)$--data from reversible magnetization. }
		\label{fig:JPR0-and-gamma}
		\vspace{-5mm}
\end{figure}

\noindent where $J^{(c)}_{m}(B,T) = J^{(c)}_{m}(0,T) \langle \cos(\phi_{n,n+1})\rangle$
is the maximum Josephson current, $s=1.5$~nm is the interlayer 
spacing, $\epsilon_{r}$ the high-frequency relative
dielectric constant and $\Phi_{0}$ the flux quantum.
Using $\omega_{pl}(0,T)Ê= c / \lambda_{L,c}(T)\sqrt\epsilon_{r}$ 
and  $\epsilon_{r}$=11.5 \cite{Gaifullin01}, we obtain the 
London penetration depth for currents along the $c$-axis, 
$\lambda_{L,c}(T)$. 
When divided by the $\lambda_{L,ab}(T)$--data from reversible 
magnetization, this yields, without any model assumptions,
the anisotropy parameter $\gamma(T) \equiv 
\lambda_{L,c}(T)/\lambda_{L,ab}(T)$, shown in the Inset to 
Fig.~\ref{fig:JPR0-and-gamma}. Typically, at $T = 0.5 T_{c}$, 
$\lambda_{L,c}\approx 240~\mu$m and 
$\lambda_{L,ab}\approx 400$ nm, so that $\gamma \approx 600$,
consistent with other data for the same material \cite{Gaifullin00}. 
Note that  $\gamma$ decreases as function of temperature. 

To analyze our JPR data in non-zero magnetic fields, we should 
divide $\omega_{pl}(B,T)$ by the zero--field result depicted in 
Fig.~\ref{fig:JPR0-and-gamma}. In the absence of a model that satisfactorily
describes $\omega_{pl}(0,T)$ over the whole temperature range, we 
resort to a spline fit to the experimental data. 
Next, we extract the vortex wandering length $r_{w}$ as follows. 
In the single vortex regime, at very low fields 
$B<B_{J}= \Phi _{0}/\lambda_{J}^{2}$, 
$B < B_{\lambda}= \Phi _{0}/4\pi \lambda_{L,ab}^{2}$, 
Bulaevskii and Koshelev derived \cite{Bulaevskii00,Koshelev00}

\begin{equation}
1 - \frac {\omega _{pl}^{2} (B,T)}{\omega _{pl}^{2} (0,T)} 
\approx  \frac {\pi B}{2 \Phi_{0}} r_{w}^{2} \ln \frac {\lambda _{J}}{r_{w}}
\label{eq:rw-Koshelev}
\end{equation}

\noindent where the Josephson length $\lambda _{J} = \gamma s$.
We stress that this relation is meaningful only for small excursions 
$r_{w} \le 0.6\lambda_{J}$, {\em i.e.} for $\langle 
\cos(\phi_{n,n+1})\rangle = \omega _{pl}^{2} (B,T)/\omega 
_{pl}^{2} (0,T) \lesssim 1$. 
More generally, one expects an increase of $1 - \langle 
\cos(\phi_{n,n+1})\rangle$ with $r_{w}$ up to a plateau for large 
$r_{w}$, as was found in recent simulations of the  evolution of $1-\langle 
\cos(\phi_{n,n+1})\rangle$ versus $\langle u \rangle/a_{0} \sim 
r_{w}/a_{0}$ for a pancake gas ($a_{0} = \sqrt{\Phi_{0} / B}$ is the intervortex 
spacing) \cite{Brandt02}. 
The numerical data show that  $1-\langle \cos(\phi_{n,n+1})\rangle$ is almost quadratic in $r_{w}$ for 
$0 \lesssim 1-\langle \cos(\phi_{n,n+1})\rangle \lesssim 0.7-0.8$, in agreement with 
Eq.~(\ref{eq:rw-Koshelev}), if the weak logarithmic dependence on $\lambda_{J}/r_{w}$ is 
disregarded. Thus, we use 

\begin{equation}
r_{w}^{2}=\frac {2 \Phi_{0}}{\pi B} (1-\langle \cos(\phi_{n,n+1})\rangle)  
\label{eq:rw-Sylvain}
\end{equation}

\noindent to obtain an approximation of the wandering length.
Since  $r_{w}= \langle( {\bf u}_{n+1} - {\bf 
u}_{n})^{2}\rangle^{1/2} = [2(u^{2}-\langle{\bf 
u}_{n}\cdot{\bf u}_{n+1}\rangle)]^{1/2}$, one has, in the case 
of completely uncorrelated layers ({\em e.g.} for a pancake 
gas), $r_{w} = \langle 2{\bf u}_{n}^{2}\rangle^{1/2} \equiv \sqrt{2}u$.
Disregarding the ``anticorrelated'' situation with 
${\bf u}_{n} \cdot {\bf u}_{n+1} < 0$, correlations between pancake 
positions in different layers yield $r_{w} < \sqrt{2}u$, {\em i.e.}, 
$r_{w}$ is a lower limit for the root mean squared (RMS)
displacement $u$ of the vortex line. 


Figure~4 shows $1-\omega _{pl}^{2} (B,T)/\omega 
_{pl}^{2} (0,T)  = 1-\langle \cos(\phi_{n,n+1})\rangle$ as 
function of temperature in different dc fields. The temperature 
dependence of the wandering length $r_{w}$, obtained  by applying 
Eq.~(\ref{eq:rw-Sylvain}), is represented in Fig.~5. For every field, 
we observe an increase of $r_{w}$ with~$T$. At constant temperature, 
$r_{w}$ increases with magnetic field,
indicating a $\omega _{pl}(B)$--dependence that deviates from the expected linear 
behavior of Eq.~(\ref{eq:rw-Koshelev}).
Another interesting feature of the $r_{w}$($T$) curve is the break in the slope 
which appears at a field-dependent temperature close to the FOT
and above which all the $r_{w}$
curves merge into one. Alternatively, one may plot the same values of 
$r_{w}$ vs $T/T_{FOT}$, where $T_{FOT}$ is the FOT temperature 
(Fig.~5b). Here, two regimes appear clearly, showing the correlation 
of the results with the FOT. For $T < 0.96T_{FOT}$, $r_{w}(T/T_{FOT})$ roughly 
overlaps for all fields, whereas for $T>0.96T_{FOT}$ the curves deviate 
from each other.

\begin{figure} [t]
		\centerline{\epsfxsize 7.0cm \epsfbox{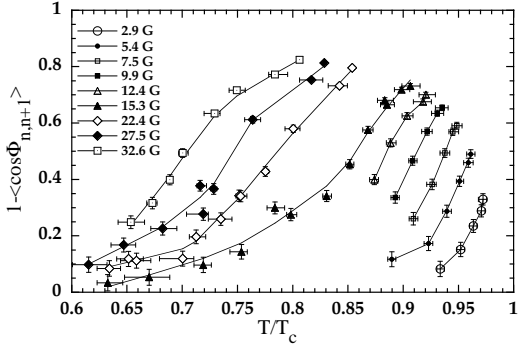}}
\caption{  $1-\langle \cos(\phi_{n,n+1})\rangle$ vs temperature for 
different magnetic fields. We extract $r_{w}$ from 
these data using Eq.(\ref{eq:rw-Sylvain}).}
\end{figure}

\begin{figure}[t]
		\centerline{\epsfxsize 7.5cm \epsfbox{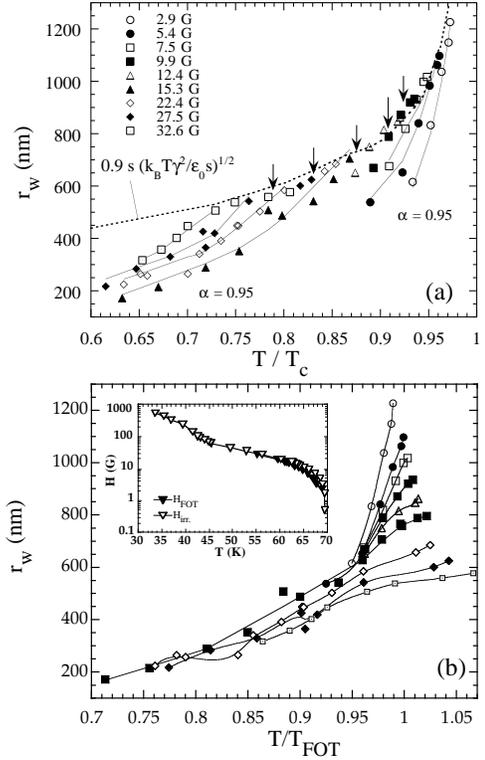}}
		\caption{ (a) : Experimental $r_{w}$ vs $T/T_{c}$ for 
		different magnetic fields in strongly underdoped BSCCO. 
		For $B=$27.5, 22.4, 15.3, 12.4 and 9.9~G, arrows show 
		the temperature of the FOT. The dotted line shows the evolution of  
		$0.9 s (k_{B}T\gamma^{2}/\varepsilon_{0}s)^{1/2}$. Solid 
		lines are fits to Eq.~(\protect\ref{eq:rw-T-Bdependence}) with the single 
		parameter $\alpha = 0.95$, omitting the term 
		$4/\pi\left(x^{2}  + \frac{1}{4} \right)$ for $B > 15$ G.
		(b) :  $r_{w}$ vs $T/T_{FOT}$. Solid 
		lines are guides to the eye. Inset : phase diagram of a sample 
		cut from the same crystal. Full and open triangles stand for the 
		FOT and the irreversibility fields, respectively.}
		\label{fig:rw-vs-t}
		\vspace{-3mm}
	\end{figure}

We now discuss the temperature and field dependence of $r_{w}$ in the 
vortex solid. The RMS  thermal vortex displacement $u$
can be obtained by equipartition of the associated elastic energy  
with the thermal energy, $U_{el} = c_{44}a_{0}^{2}(u^{2}/s) = k_{B}T$. 
The vortex lattice tilt modulus 

\begin{eqnarray}
	c_{44}({\bf k})   &\approx &  \frac {B^{2}/\mu_{0}}{1+\lambda_{c}^{2}k_{\parallel}^{2}+\lambda_{ab}^{2}Q_{z}^{2}}
	+\frac{\varepsilon_{0}}{2\gamma^{2}a_{0}^{2}}\ln 
	\left[\frac{k_{max}^{2}}{K_{0}^{2}+(Q_{z}/\gamma)^{2}}\right] \nonumber \\
	& &+\frac {\varepsilon_{0}}{2\lambda_{ab}^{2}Q_{z}^{2}a_{0}^{2}}\ln 
	\left(1+\frac{a_{0}^{2}}{21.3r_{w}^{2}}\right), 
	\label{eq:c44}
\end{eqnarray}

\noindent  calculated by Koshelev and Vinokur \cite{Koshelev98}
and Goldin and Horowitz \cite{Goldin98}, consists of three terms: the 
nonlocal collective (lattice) term, the vortex line tension term, determined by 
Josephson coupling between layers, and a third term due to the
electromagnetic dipole interaction between pancakes. Of 
particular interest here is the logarithmic correction to the 
temperature dependence of the second term, introduced by the cutoff
$k_{max} = \pi/r_{w}$, which corresponds to the smallest
meaningful deformation \cite{Koshelev98,Goldin98}. To 
proceed, we evaluate $U_{el}$ at the typical vortex line deformation wavevectors
parallel and perpendicular to the layers, $k_{\parallel} \approx \pi/r_{w}$ 
and $Q_{z} \approx \pi\gamma/2a_{0} \ll 2\pi/s$. Writing $K_{0} = 
\sqrt{4\pi}/a_{0}$, $r_{w}^{2} = \alpha u^{2}$ and $x = a_{0}/r_{w}$, 
equipartition yields

\begin{eqnarray}
	r_{w}^{2} & \approx  & \alpha s^{2}	\frac{k_{B}T\gamma^{2}}{\varepsilon_{0}s} 
	                       \left[   \frac{4}{ \pi \left(x^{2}  + \frac{1}{4} \right) }  + 
						            \frac{1}{2} \ln\left( 0.66 x^{2} \right) \right. 
									\nonumber \\
	 & &                   \left. + \frac{2}{\pi^{2}} \left( \frac{a_{0}}{\lambda_{L,ab}} \right)^{2}
			                        \ln \left( 1+\frac{x^{2}}{21.3} \right) 
			                        \right]^{-1}.
	\label{eq:rw-T-Bdependence}
\end{eqnarray}

\noindent All parameters in Eq.~(\ref{eq:rw-T-Bdependence}), and notably 
$\varepsilon_{0}(T)/\gamma^{2}(T) = 
\Phi_{0}^{2}/4\pi\mu_{0}\lambda_{L,c}^{2}$, are known from experiment, 
which allows a direct comparison to the $r_{w}(T)$--data. Very good
agreement is obtained for the magnitude, the temperature, as well as 
the field dependence of  $r_{w}$  for the lowest three fields, using 
the single free parameter $\alpha = 0.95$. For higher fields, 
Eq.~(\ref{eq:rw-T-Bdependence}) gives the correct magnitude of $r_{w}$, 
but too weak a temperature dependence. However, excellent fits of both 
the temperature and field dependence can be obtained for all fields,
with the same $\alpha = 0.95$, by omitting the nonlocal collective term, 
[{\em i.e.} $4/\pi\left(x^{2}  + \frac{1}{4} \right)$ in 
Eq.~(\ref{eq:rw-T-Bdependence})], see Fig.~5a.   While the
main temperature dependence  of $r_{w}$ comes from the prefactor
$\gamma^{2}T/\varepsilon_{0}$ in Eq.~(\ref{eq:rw-T-Bdependence}) (dotted line in 
Fig.~\ref{fig:rw-vs-t}), the behavior of $r_{w}$ in the vortex solid can only be understood 
as the result of the logarithmic correction arising from the softening 
of the line tension term by thermal fluctuations 
\cite{Goldin98,Koshelev98}. The field dependence, 
originating from $Q_{z}$, explicitly indicates that vortex lines are 
correlated (line-like) on distances that well exceed the layer spacing $s$.

The experimental data can also be used to compare the terms
entering Eq.~(\ref{eq:c44}). Deep inside the vortex solid, the line 
tension always dominates over the magnetic coupling and the nonlocal 
collective contribution. At very low fields ($B < 10$ G), the line 
tension term is largest all the way 
to the FOT. Eq.~(\ref{eq:rw-T-Bdependence}) then reduces to Eq.~(40) 
of Ref.~\onlinecite{Goldin98} with $Q_{z} \approx 
\pi\gamma /2 a_{0}$ instead of $2\pi/s$. At higher fields, the nonlocal 
collective contribution is expected to increase, 
eventually exceeding the Josephson coupling (line tension) term close to the 
FOT (for $B > 20$ G).  Nevertheless, the very good fits obtained when 
\em only \rm the line tension term is taken into account in 
Eq.~(\ref{eq:rw-T-Bdependence}) suggest that 
the line tension term always dominates $c_{44}$ near the FOT. 
Moreover, we find that the electromagnetic coupling as well as the 
shear contribution to $U_{el}$ are, under all 
circumstances, negligible. This renders Lindemann-like 
\cite{Blatter96} or dislocation-mediated (Kosterlitz-Thouless like) melting, 
as well as the vortex--line evaporation \cite{Dodgson00} scenarios very unlikely.
Rather, the large thermal excursions of pancake vortices bring about the 
softening of the line tension contribution to $c_{44}$ for the 
large-wavevector modes that lead to the FOT. This would comply with recent 
measurements showing that vortex lattice order is not a prerequisite for 
the FOT \cite{porous}. For deformations with 
smaller wavevectors, Josephson coupling still contributes to 
the line tension even in the vortex liquid, leading to, {\em e.g.}, 
the anisotropic vortex response to columnar defects in heavy-ion 
irradiated samples.

Summarizing, we have carried out JPR measurements on heavily underdoped 
Bi$_{2}$Sr$_{2}$CaCu$_{2}$O$_{8}$ crystals.
These data yield the $c$--axis penetration depth, the anisotropy parameter $\gamma(T)$, 
and the wandering length of vortex lines $r_{w}$. The observed temperature and field 
dependences of $r_{w}$ suggest that thermal fluctuations soften
the Josephson coupling contribution to the tilt modulus for short 
wavelengths \cite{Goldin98}, which leads us to 
believe that this softening drives the FOT.

We thank the Europe Scientific Foundation VORTEX program 
and the Nederlandse Organisatie voor Wetenschappelijk Onderzoek
for financial support. PG acknowledges support of the Polish 
Government, grant PBZ-KBN-013/T08/19.


\end{document}